\documentclass[onecollarge,runningheads]{svjour2}
\smartqed
\usepackage{graphicx}
\newcommand{\etal}{et al.}
\newcommand{\apj}{ApJ}

\newcommand{\mnras}{MNRAS}
\newcommand{\pasp}{PASP}
\journalname{Experimental Astronomy}
\begin{document}

\title{The Molecular Hydrogen  Explorer H2EX}

%\titlerunning{The H$_2$ Explorer space mission H2EX}

\author{F. Boulanger \and J. P. Maillard \and P. Appleton \and E. Falgarone
\and G. Lagache \and B. Schulz \and B. P. Wakker \\ % \and
 A. Bressan \and J. Cernicharo \and V. Charmandaris \and L. Drissen \and G. Helou 
\and T. Henning \and T. L. Lim \and E. A. Valentjin \\ % \and
 A. Abergel \and J. Le Bourlot \and M. Bouzit \and S. Cabrit\and F. Combes \and
J. M. Deharveng \and P. Desmet \and H. Dole \and C. Dumesnil \and A. Dutrey
\and J. J. Fourmond \and E. Gavila \and R. Grang\'e \and C. Gry \and
P. Guillard \and S. Guilloteau \and E. Habart \and B. Huet\and   C. Joblin \and M. Langer
\and Y. Longval \and S. C. Madden \and C. Martin \and
M. A. Miville-Desch\^enes \and G. Pineau des For\^ets \and  E. Pointecouteau 
\and H. Roussel \and L. Tresse \and L. Verstraete \and F. Viallefond 
\and F. Bertoldi \and J. Jorgensen 
\and J. Bouwman \and A. Carmona \and O. Krause 
 \and A. Baruffolo \and C. Bonoli \and F. Bortoletto \and L. Danese \and
G. L. Granato \and C. Pernechele \and R. Rampazzo \and L. Silva \and G. de Zotti   
  \and  J. Pardo 
 \and M. Spaans \and F. F. S. van der Tak \and W. Wild
\and M. J. Ferlet   \and S. K. Ramsay Howat \and M. D. Smith \and  B. Swinyard 
\and G. S. Wright 
 \and J. Joncas \and P. G. Martin 
   \and C. J. Davis \and  B. T. Draine
\and P. F. Goldsmith \and A. K. Mainzer \and P. Ogle \and S. A. Rinehart 
 \and G. J. Stacey \and A. G. G. M. Tielens   
\footnote{The coauthor names are grouped by partner institute and the
  institutes by country.}}

\authorrunning{Boulanger, Maillard \etal}

\institute{F. Boulanger (H2EX Principal Investigator), A. Abergel,
  M. Bouzit, H. Dole, C. Dumesnil, J. J. Fourmond, E. Habart, G. Lagache,
  M. Langer, Y. Longval, M. A. Miville-Desch\^enes, G. Pineau des
  For\^ets, L. Verstraete \at Institut d'Astrophysique Spatiale, CNRS \&
  Universit\'e Paris-Sud, B\^at 120 - 121, 91405 Orsay, France\\
  \email{francois.boulanger@ias.u-psud.fr}\\ 
\and J. P. Maillard (Payload Scientist), H. Roussel 
\at Institut d'Astrophysique de Paris, CNRS et Universit\'e P. \& M. Curie, 
98\,bis Blvd Arago, 75014 Paris, France\\
  \email{maillard@iap.fr}\\ 
\and S. Cabrit, F. Combes, E. Falgarone,  F. Viallefond (LERMA), J. Le
  Bourlot (LUTH) \at Observatoire de Paris, 75014 Paris, France\\ 
\and P. Desmet, E. Gavila, B. Huet \at Thales Alenia Space,
  06322 Cannes La Bocca, France\\ 
\and A. Dutrey, S. Guilloteau \at
  Observatoire de Bordeaux, 33270 Floirac, France\\ 
\and C. Joblin, E. Pointecouteau \at Centre d'Etudes Spatiales des Rayonnements, 31028
  Toulouse, France\\ 
\and J. M. Deharveng, R. Grang\'e, C. Gry, L. Tresse
  \at Laboratoire d'Astrophysique de Marseille, 13376 Marseille, France\\
\and S. C. Madden, C. Martin \at Service d'Astrophysique, B\^at. 609,
  Orme des Merisers, CEA Saclay, 91191 Gif-sur-Yvette, France\\ 
\and  F. Bertoldi, J. Jorgensen \at Argelander Institute for Astronomy,
  University of Bonn, Auf dem H\"ugel 71, 53121 Bonn, Germany\\ 
\and J. Bouwman, A. Carmona, O. Krause, T. Henning \at Max-Planck-Institut
  f\"ur Astronomie, K\"onigstuhl 17, 69117 Heidelberg, Germany\\ 
\and V. Charmandaris \at University of Crete, Department of Physics, P.O. Box
  2208, 71003 Heraklion, Greece\\ 
\and A. Bressan, A. Baruffalo, C. Bonoli, F. Borletto, G. L. Granato,
  R. Rampazzo, G. de Zotti, \at
  INAF-Osservatorio Astronomico di Padova, Vicolo dell'Osservatorio 5,
  35122 Padova, Italy\\ 
\and L. Danese \at Scuola Int. Sup. di Studi
  Avanzati (SISSA), Via Beirut 4, 34014 Trieste, Italy\\ 
\and C. Pernechele
  \at INAF-Osservatorio Astronomico di Cagliari, Poggio dei Pini Strada 54,
  09012 Capoterra (CA), Italy\\ 
\and L. Silva \at INAF-Osservatorio
  Astronomico di Trieste, Via Tiepolo 11, 34143 Trieste, Italy\\ 
\and J. Cernicharo, J. Pardo \at Dep. de Astrofis. Mol. e Infrarroja, Inst. de
  Estructura de la Materia, CSIC, Serrano 121, 28006 Madrid, Spain\\ 
\and E. A. Valentjin, M. Spaans, F. F. S. van der Tak, W. Wild \at Netherlands
  Institute for Space Research (SRON), Groningen, The Netherlands\\ 
\and M. J. Ferlet, T. L. Lim, B. Swinyard \at Rutherford Appleton Laboratory,
  Chilton, Didcot OX11 0QX, United Kingdom\\ 
\and S. K. Ramsay Howat, G. S. Wright \at UK Astronomy Technology Center, 
Royal Observatory, Edinburgh EH9 3HJ, United Kingdom\\ 
\and M. D. Smith \at Centre for Astrophys. and Planet. Science, School of
  Physical Sciences, The University of Kent, Canterbury CT2 7NH, United Kingdom\\ 
\and L. Drissen, G. Joncas, \at D\'epartement de Physique, de G\'enie Physique et
  d'Optique and Observatoire du mont M\'egantic, Universit\'e Laval,
  Qu\'ebec, QC G1K 7P4, Canada\\ 
\and P. G. Martin \at Canadian Institute for Theoretical Astrophysics, 60
  St-George st, Toronto, Ontario, M5S 3H8, Canada\\ 
\and P. Appleton, G. Helou, P. Ogle, B. Schulz \at IPAC,
  Cal. Institute of Technology, Mail Code 100-22, 770 South Wilson Avenue,
  Pasadena, CA 91125, USA\\
\and C. J. Davis  \at Joint Astronomy Centre, Hilo, HI 96720, USA\\  
\and B. T. Draine \at Princeton University
  Observatory, Princeton, NJ 08544, USA\\ 
\and P. F. Goldsmith, A. K. Mainzer \at Jet Propulsion Laboratory, 4800 Oak Grove Drive,
  Pasadena, CA 91109, USA\\ 
\and S.A. Rinehart \at NASA Goddard Space
  Flight Center, Greenbelt, MD 20771, USA \\ 
\and A. G. G. M. Tielens \at NASA Ames
  Research Center, Moffett Field, CA 94035, USA\\ \and G. J. Stacey \at
  Department of Astronomy, Cornell University, Ithaca, NY 14853, USA\\ 
\and B. P. Wakker \at Department of Astronomy, University of Wisconsin,
  Madison, WI 53706, USA\\ } 
\date{Received: date / Accepted: date}
  \maketitle

\begin{abstract}
The Molecular Hydrogen Explorer, $H2EX$, was proposed in
response to the ESA 2015 - 2025 Cosmic Vision Call as a medium class space
mission with NASA and CSA participations. The mission, conceived to
understand the formation of galaxies, stars and planets from molecular
hydrogen, is designed to observe the first rotational lines of the
H$_2$ molecule (28.2, 17.0, 12.3 and 9.7\,$\mu$m) over a wide field, and
at high spectral resolution.  $H2EX$ can provide an inventory of warm
($\geq$ 100\,K) molecular gas in a broad variety of objects, including
nearby young star clusters, galactic molecular clouds,
active galactic nuclei, local and distant galaxies. The rich array of
molecular, atomic and ionic lines, as well as solid state features
available in the 8 to 29\,$\mu$m spectral range brings additional
science dimensions to $H2EX$.\\ 
We present the optical and mechanical
design of the $H2EX$ payload based on an innovative Imaging Fourier Transform
Spectrometer (IFTS) fed by a 1.2\,m telescope. The 20\rq$\times$20\rq\
field of view is imaged on two 1024$\times$1024 Si:As detectors. The
maximum resolution of 0.032\,cm$^{-1}$ (FWHM) means a velocity
resolution of 10\,km\,s$^{-1}$ for the 0\,--\,0 S(3) line at
9.7\,$\mu$m. This instrument offers the large field of view necessary to
survey extended emission in the Galaxy and local Universe galaxies as
well as to perform unbiased extragalactic and circumstellar disks
surveys.  The high spectral resolution makes $H2EX$ uniquely suited to
study the dynamics of H$_2$ in all these environments.  The mission plan
is made of seven wide-field spectro-imaging legacy programs, from the
cosmic web to galactic young star clusters, within a nominal two years
mission. The payload has been designed to re-use the $Planck$ platform
and passive cooling design.
\end{abstract}

%%-----------------------------

\section{Introduction}

The dynamics and energetics of molecular gas link the formation of
galaxies, stars and giant planets within astrophysics.  Progress in these
areas depends critically on our ability to survey the molecular gas content
of the Universe, and the energy radiated and carried away by its
energetically processed fraction.  Present studies primarily rely on
tracing molecular gas through the observation of low rotational emission
lines of CO. This provides a selective, if not biased, view of the
molecular Universe.  CO observations may well miss a major fraction of the
molecular gas in galaxies hidden in their low metallicity outskirts as well
as key aspects of the energetic interplay between galaxies and the
intergalactic medium, and between stars and their nascent clouds.
Molecules are known to be ubiquitous. They have been observed in both the
local and the distant Universe, in environments as diverse as
proto-planetary disks, star forming regions, the diffuse interstellar
medium, galactic disks, cooling flows, and distant Ultra-Luminous Infrared
Galaxies (ULIRGs).  However, observations have so far remained blind to the
main constituent of this gas, H$_2$, leaving its role in cosmic evolution
and its diagnostic power, mostly unexplored.

The $Copernicus$ and $FUSE$ satellite observations studied the electronic
transitions in the UV and were sensitive to very low H$_2$ column densities,
but the UV absorption by dust is also very important and these observations
are limited to tenuous regions. They miss H$_2$ in shielded regions and
thus, most of the H$_2$ in the Universe. The near-infrared
rovibrational transitions (1\,--\,0 transitions and above) are routinely
imaged from ground-based telescopes. Emission in these lines mostly results
from fluorescence following UV absorption. Very high densities and
temperatures are required for the collisional excitation of the
vibrationally excited levels.  The first rotational
transitions in the mid-infrared (0\,--\,0 S(0) at 28.2\,$\mu$m, S(1) at
17.0\,$\mu$m, S(2) at 12.3\,$\mu$m, S(3) at 9.7\,$\mu$m...) are more
relevant to baryonic structure evolution because they relate to the
energetics of the bulk of the molecular gas.  Most of the H$_2$ is too cold
to emit strongly, but UV pumping and the dissipation of both ordered,
and turbulent kinetic energy give rise to significant emission in the pure
rotational lines.

The Earth's atmosphere is mostly opaque to several of the lowest
ground-state lines of H$_2$. For the transitions occurring at
wavelengths where the atmosphere is reasonably transparent, the thermal
background from the warm telescope, spectrometer and sky, limits the
sensitivity.  To change conclusively this situation and make possible
the direct observation of all astrophysical environments where molecular
hydrogen is present and significant to the gas energetics, we designed a
dedicated infrared space mission, the \textit{Molecular Hydrogen
Explorer} or $H2EX$, associating a wide field of view
(20\rq$\times$20\rq) and a high spectral resolution (up to $3\times
10^4$) for spectral imaging. This project takes advantage of
technological development in lightweight mirrors, panoramic infrared
detectors, passive cooling and cryogenic coolers, to image
the spatial and velocity distribution of molecular gas through
observations of the lowest rotational transition lines of H$_2$. The
rich array of molecular, atomic and ionic lines, as well as solid state
features available in the 8 to 29\,$\mu$m spectral range brings
additional science dimensions to $H2EX$. 
%Its uniqueness compared to
%present and planned infrared facilities comes from the combination of
%wide-field imaging capability with versatile spectral resolution
%resolving power.

$H2EX$ was submitted to ESA in answer to the Cosmic Vision 2015-2025
call as a medium class space mission but was not selected for a Phase~A
study.  The science program, which still remains of actuality, is
presented in Sect.~\ref{sec:science}. It translates into several legacy
surveys described in Sect.~\ref{sec:surveys}.  They define the project
science requirements (Sect.~\ref{requir}).  The optical and mechanical
design of the payload based on an innovative Imaging FTS are described
in Sects.~\ref{h2ex_solution} to \ref{h2ex_meca}. The spacecraft and
mission profile are oulined in Sect.~\ref{spacecraft}. Finally, to show
its originality, $H2EX$ is replaced in the context of the other infrared
facilities on ground and in space (Sect.~\ref{h2ex_context}).

\section{$H2EX$ Science Program}
\label{sec:science}

$H2EX$ is designed to address five key questions related to the
formation of galaxies, stars and planets, and the chemical evolution of
matter in space that are briefly described below.

     \subsection{How is the luminous Universe taking shape?}
\begin{figure}[!ht]  % fig. 1
\centering
\includegraphics[width=0.85\textwidth]{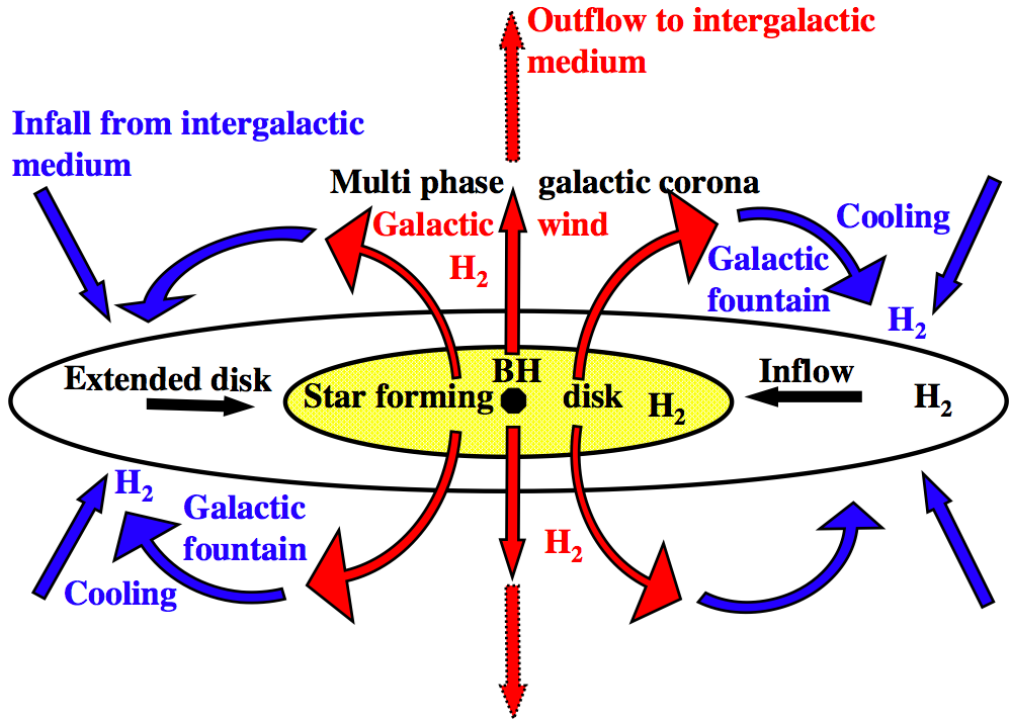}
%/home/mure1/maillard/instrum/H2EX/proposal_CV/galaxy_cartoon.ps}
\caption{The gas disk of the galaxy is fed by infall from intergalactic
space and by gas raining down from the halo.  This gas flows inwards and
fuels star formation as well as the activity of the central black hole.  The
arrows outline these inflows and outflows, and the gas cooling and
heating. H$_2$ is a key actor in all stages. }
\label{fig:infall_outflow} 
  \end{figure}

Understanding galaxy formation will be a major challenge for many years to
come. Although the observational basis has been growing steadily, we have
only started to come to some understanding of the physics that gives rise
to the observed numbers, masses, morphologies and chemical composition of
galaxies as a function of cosmic environment and time.  The growth of
galaxies depends on their mass and environment and the evolution of the gas
within dark matter haloes. 
The build-up of baryonic mass in galaxies is regulated by a
complex interplay of energy and matter between different gas phases. It is
affected by the formation of stars, massive central black holes and flows
in and out of galactic disks and haloes (Fig.~\ref{fig:infall_outflow}).
H$_2$ is present in all stages because its formation is a natural outcome
of gas cooling, and because stars are made from this phase of matter.

$H2EX$ is designed to survey galactic-scale H$_2$ emission associated
with star formation, gas accretion and feedback, as well as
spectroscopic fingerprints (dust features, ionized gas lines) of a large
number of infrared luminous galaxies.  $Spitzer$ Space Telescope and
Infrared Space Observatory ($ISO$) observations have revealed a new
class of extremely luminous molecular hydrogen emission galaxies with
L(H2) = 10$^{40}$ -- $10^{43}\, {\rm erg \, s^{-1}}$ in pure rotational
molecular hydrogen emission lines but with relatively weak total IR
emission, $\rm L(H_2)/L(IR)> 10^{-3}$
\cite{rigopoulou,appleton,egami,ogle}.  In these galaxies, the S(5) line
at 6.9\,$\mu$m is observed to be comparably bright to the S(1) and S(3)
lines.  $H2EX$ can observe this line up to a redshift of 2.5. Its unique
ability to do wide-field spectroscopic surveys in the infrared places it
as a key player in the opening up of cosmic discovery space.

     \subsection{What is the H$_2$ contribution to the missing baryons?}
Galaxies dynamics and weak lensing surveys suggest that a significant
fraction of the dark matter in the Universe is in the virialized haloes
of $\rm L\sim L_*$\footnote{Break in the Schechter' galaxies luminosity
function \cite{schechter}} galaxies but the observed mass in condensed
baryons -- stars and gas -- is only about 8\% of the total
\cite{fuku_peebles04}.  Intergalactic shocks may have made some of the
baryons too hot to fall into dark matter haloes.  This interpretation is
supported by evidence that a substantial amount of baryons is in the
warm-hot intergalactic medium, with temperatures in the range
10$^5$--10$^7$\,K \cite{dave}. For galaxies like the Milky Way -- far
from massive cosmic structures -- it is more likely that much of their
baryonic mass has been blown out of their disks, possibly out of their
haloes, or may be hidden from view. $H2EX$ is designed to test two
aspects of the missing baryons problem.  (1) A survey in the local
Universe will test whether many of the missing baryons could be H$_2$
gas hiding in the extended disk and/or haloes of galaxies.  This gas
would have escaped detection through CO observations because of its low metallicity
and clumpiness.  In low-metallicity regions, the H$_2$ rotational lines
are the main pathway through which the gas cools. So a deep unbiased
search for H$_2$ emission provides a unique means to test this
possibility.  (2) $H2EX$ spectroscopic imaging of galactic winds in
local Universe starburst galaxies and active galactic nuclei will
quantify the energy radiated, and the mass and kinetic energy carried by
molecular gas.  They will complement optical to X-ray observations that
only probe the warm and hot ionized components of galactic winds.
   
     \subsection{What controls star formation efficiency?}
Star formation is the end-process of a multiscale condensation of gas
under the effect of gravity. It is controlled by the dissipation of
interstellar turbulent kinetic energy in molecular clouds which allows
gravity to take over. Once stars have formed, their feedback through
jets and winds, stabilizes their nascent cloud. Milky Way and local
group observations of H$_2$ will probe molecular cloud energetics from
their formation out of the diffuse interstellar medium to their
dispersal by massive stars. $H2EX$ will measure the rate of condensation
of diffuse matter towards gravitationally bound molecular clouds.  It
will also reveal the molecular component of star forming regions that
carries most of the momentum provided by stellar feedback.

    \subsection{How do giant planets form?}
%The detection of a rapidly growing number of known extra-solar giant
%gaseous planets plays up the importance of understanding their
%formation. 
Planets are thought to form in the circumstellar disks that surround
proto-stars and pre-main-sequence stars. H$_2$ dominates the disks mass
and dynamics. Two main scenarios are proposed for the formation of giant
planets: gas accretion onto massive rocky cores or direct formation out
of the gas as a result of gravitational instability.  In the simplest
scenario, the core accretion process requires that H$_2$ is present at
late stages. The gravitational instability mechanism predicts formation
at earlier times out of a very massive H$_2$ disk.  To constrain the
planet formation scenarios it is imperative to obtain direct
observational evidence on the disk H$_2$ content and how long it lasts.
$H2EX$ offers the required combination of field and spectral resolution
to survey young star clusters for H$_2$ emission from planet-forming
disks.

     \subsection{Where do complex organic molecules form?}
The formation of H$_2$ initiates a network of chemical reactions leading to
an impressive molecular complexity in space. Within the shielded interiors
of dense clouds, molecules stick to grain surfaces which become the sites
of further chemical evolution. Some of these grain surface molecules
become constituents of protoplanetary disks, where they contribute to the
formation of icy planetesimals such as the comets in our Solar System.  The
richness of molecular chemistry in dense gas is highlighted by observations
of large organic molecules in proto-stellar cores heated by nascent stars,
where ices return to the gas.  Evolved stars are also known to be sites for
forming large carbonaceous molecules.  The evolutionary connection between
these circumstellar species and those in the interstellar gas has yet to be
fully elucidated.\\  
The $H2EX$ spectrometer offers the possibility of charting the hidden
roots of complex molecules in space.  High resolution mid-infrared
spectroscopy is the only means by which we can detect rovibrational
lines from organic molecules (C$_n$, HC$_n$H, H$_2$C$_n$H$_2 $,
CH$_3$CH$_2$CH$_3$,...) that are the key nodes of organic
chemistry. Such species do not have permanent dipole moments and hence,
have no submillimetric rotational transitions. With its wide field of view,
$H2EX$ will also unveil spatial variations of molecular abundances and
thereby the chemical interplay between gas, ices and dust, and their
evolution over time.

\section{$H2EX$ Legacy Surveys}
\label{sec:surveys} 
With the aim of answering the astrophysical
questions reviewed above seven $H2EX$ Legacy surveys have been defined
(Table~\ref{tab:surveys}).

\begin{table}[!ht]
\caption{The $H2EX$ surveys}
\centering
\begin{tabular}{l |rrrrr}
\hline\noalign{\smallskip}
Survey Name&  Fields &Area &Sources  &Spectral~~~~~~& R~~~~\\ 
           & Number & deg$^2$& count &domain~~~~~~&$\lambda/\Delta\lambda~$ \\ 
%\hline
\tableheadseprule\noalign{\smallskip}
1) Cosmic Web I & 90  & 10  & $3\times 10^5$  & 8 - 25$\mu$m & 10$^2$   \\
2) ~~~~~--~~~~~~~~~~ II & 40  & 4 & $\sim 10^5$  &8 - 25$\mu$m & 10$^3$ \\
3) ~~~~~--~~~~~~~~~~ III& 40  & 4 & $\sim 10^5$  &S(1), S(2), S(3) & 10$^3$\\
4) Local Universe     & 100 & 11  &100    &S(0), S(1), S(2), S(3)& $10^4$ \\
5) Milky Way        & 75  & 8.3   &   &S(0), S(1), S(2), S(3)  & 1-3 $\times 10^4$ \\
6) Circumstellar disks   & 25  & 2.8   & $\sim 3000$ & S(3) & 1-3 $\times 10^4$ \\
7) Astrochemistry  & 12  &1.3  & &8 - 25$\mu$m   & $10^4$  \\
\noalign{\smallskip}\hline
\end{tabular}
  \label{tab:surveys}
\end{table}

  They consist of a dedicated series of
targeted observations, each producing data cubes which combine spatial
 and spectral information with a field of view and
spectral resolution appropriate for the science goal.  Long integration
times (24 to 100\,hours/field) to obtain the best sensitivity are
planned.

\paragraph{The Cosmic Web surveys} 
include three wide-field extragalactic surveys with distinct
combinations of spectral band and spectral resolution.  The low resolution
survey (I) is optimized to obtain spectra of dust emission and measure
high equivalent width emission lines from the gas.  The (II) and (III) medium spectral
resolution surveys  are optimized to detect H$_2$ and ionic fine
structure lines.  The wide-band filter survey (II) covers all redshifts.
The narrow bands survey (III) enables optimal sensitivity targeting
for specific Cosmic Web structures at selected redshifts.

\begin{figure}[!ht] % Fig. 2
\centering
\includegraphics[width=0.75\textwidth]{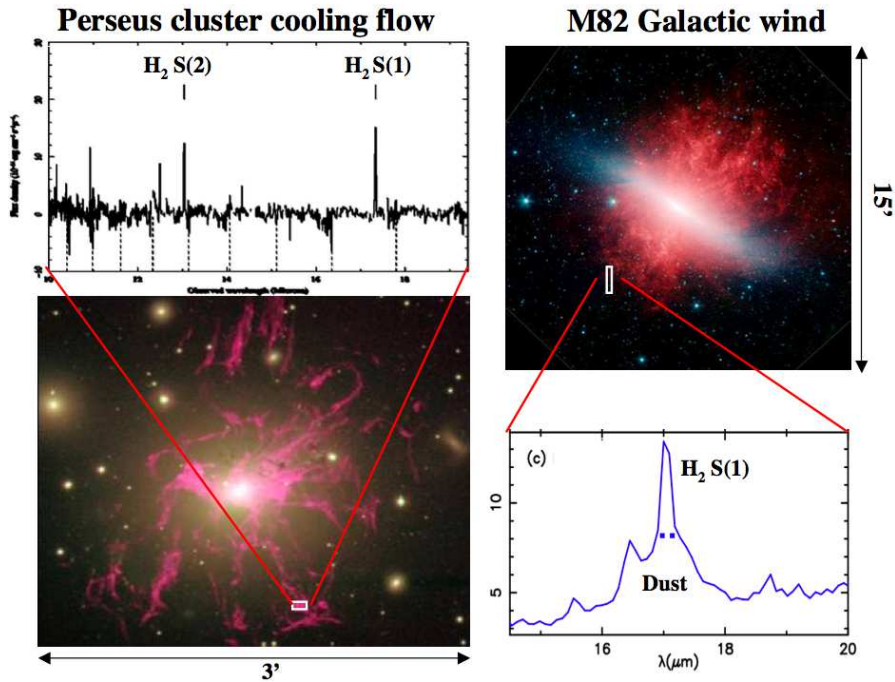}
%/home/mure1/maillard/instrum/H2EX/proposal_CV/Figure2.ps}
\caption{The Perseus cooling flow and the M82 galactic wind illustrate
the presence of H$_2$ in the gas that flows in and out of galaxies.  In
the Perseus A cooling flow optical image the network of filaments have
bright H$_2$ emission \cite{johnstone}.  In the M82 starburst galaxy
\cite{engelbracht}, the wind is observed to be loaded with molecular gas
entrained and/or formed in the interaction of the hot wind with galactic
halo gas. In both cases, the small rectangles show the very small
fraction of the galactic halo observed by $Spitzer$. }
\label{fig:composite}
\end{figure}

\paragraph{The Local Universe survey} 
comprises observations of 100 extragalactic sources sampling galaxy
types and extragalactic environments, including active galactic nuclei,
interacting galaxies, groups and nearby clusters with cooling flows
(e.g. the Perseus A cooling flow and the M82 starburst galaxy in
Fig.~\ref{fig:composite}).  H$_2$ will be observed over the whole extent
of the sources, including galaxies extended disk and haloes with a
spectral resolution of $10^4$ to measure the gas velocity. Such a deep
search for low brightness molecular gas in galaxies would have a
major impact on our understanding of the dynamical and chemical evolution
of galaxies and star formation. 
%The discovery of molecular gas in the extended
%disks of spirals would be a highlight of the mission.

\paragraph{The Milky Way survey} will include the full diversity of galactic
environments, diffuse ISM, giant molecular clouds with
photo-dissociation regions and shocks around massive YSOs, supernovae
remnants, planetary nebulae, the Galactic Center. Each of these regions
offers a specific class of problem due to the various natures of the
sources of excitation. Detailed studies of H$_2$ in galactic
environments will ground the interpretation of $H2EX$ observations of
distant galaxies.  The observations will be performed with the highest
spectral resolution to resolve the H$_2$ line profiles, making possible
to reach the turbulence conditions and the global kinematics of the gas
in all these environments. Ionic lines present in the same filter
bandpasses will be a complementary indication on the physical conditions
in different parts of each field.

\paragraph{The Circumstellar Disks survey} will target clusters within a few 100\,pc
from the Sun, spanning a range of ages from 1 to several 10\,Myr.  These
measurements will constrain the age at which gas giant planets are
forming. They will be performed with the highest spectral
resolution to obtain the needed spectral contrast to detect H$_2$
rotational line emission from proto-planetary disks 
and to measure the line width and thereby, the Keplerian radius at which
the emission originates. The field of view matches the angular extent of
nearby star clusters. It will permit the simultaneous observation of a
large ($\geq 100$ per field) number of young stellar objects in various
stages of their evolution.  The large number of objects will allow us to capture
the short evolutionary time span when the inner, planet forming part of the
disk becomes optically thin in the dust, but gas accretion on planets has
yet to occur. 

\paragraph{The Astrochemistry survey} will use the large $H2EX$ spectral
 coverage to detect minor gaseous molecular constituents, organic
 molecules that lack microwave rotational transitions, and ices, like
 CO$_2$, by their absorption signatures against bright extended
 sources. Observations of stellar clusters will provide unprecedented
 statistics on solid matter in young stellar objects.  The heavier
 isotope, HD has its higher J\=\,4-3, 5-4 and 6-5 lines occurring at 28.5,
 23.0 and 19.4 µm. These line have been observed with $ISO$ and
 $Spitzer$ by \cite{Bertoldi99} and \cite{Neufeld06} in Galactic star
 forming regions and the IC~443 supernova remnant. Two of these HD lines
 are within the spectral range of the S(0) and S(1) of H$_2$
 observations. HD detections will be used in combination with H$_2$ data
 to measure the abundance ratio [HD/H$_2$] across the Galaxy and in
 distant galaxies.

\section{The $H2EX$ science requirements}
\label{requir}
Through the legacy surveys  the payload design and
 operations of $H2EX$ are driven by the following science requirements:

\begin{figure}[!ht] % fig. 3
\centering
\includegraphics[width=0.65\textwidth]{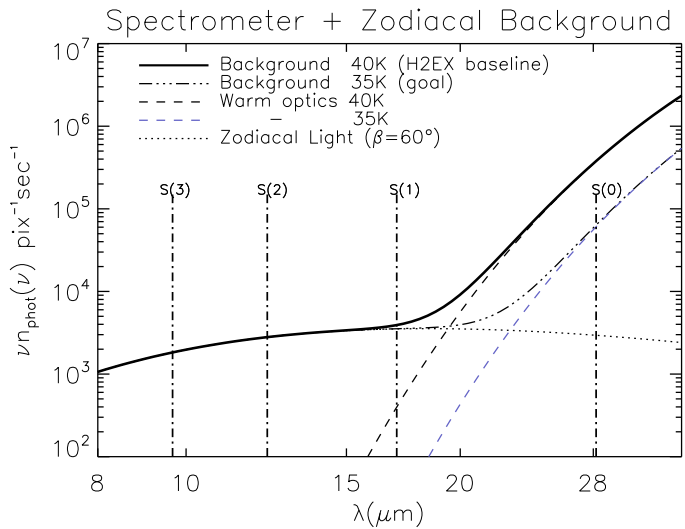}
%/home/mure1/maillard/instrum/H2EX/proposal_CV/bg35_40.ps} 
\caption{Total photon flux per frequency octave as a function of wavelength
and warm optics temperature for the instrument parameters given in 
Table~\ref{tab:detspecs}, with the position of the four H$_2$ lines  marked. The
zodiacal background is shown for an ecliptic latitude $\beta = 60^\circ$ and 
a solar elongation of $90^\circ$.}
\label{fig:background}
 \end{figure} 

\paragraph{1.}  The detection of the first pure rotational emission lines 0\,--\,0
S(0), S(1), S(2) and S(3) of H$_2$ imposes a wavelength coverage at least
from 9 to 29\,$\mu$m.
\paragraph{2.} The program must be conducted from space because only
S(1) at 17.0\,$\mu$m, and S(2) at 12.3\,$\mu$m are reasonably
accessible, from a very good infrared site of high altitude.  In any
case, the sensitivity for the rotational H$_2$ lines observable from
ground is limited by the strong thermal background from the sky and the
telescope at these wavelengths. For an optimum sensitivity in the
mid-infrared, the solution prepared for $JWST$ and also for $Planck$ and
$Herschel$, consists of placing the spacecraft at the L2 point of the
Sun-Earth system, to allow zodiacal-light-limited performance. However,
for an equilibrium temperature of 40\,K the thermal background from the
payload becomes prominent beyond 20\,$\mu$m
(Fig.~\ref{fig:background}). A temperature of 35\,K would be preferable,
and is considered as a goal.
\paragraph{3.} The study of the complete distribution of H$_2$ in 
extended regions like the intergalactic medium of galaxy clusters, in
the halo of nearby galaxies, in giant molecular clouds, calls for an
instrument able to map a wide field to perform unbiased surveys.  A
value between 10 and 30\rq\ field seems a good compromise. Thus, a
wide-field \textbf{integral field spectrometer} (IFS) is the most suited
instrument, which from the spectroscopy in all points of a field makes
possible imagery in specific spectral lines.
\paragraph{4.} Only a spectroscopic method of detection of the emission
lines makes possible to separate with a high contrast the lines from the
continuum, to resolve the line profiles making possible to derive the
kinematics, multiple velocity components and the physical
conditions.  The line widths to measure can go from a few km\,s$^{-1}$
in the local interstellar medium to several hundreds of km\,s$^{-1}$ in
external galaxies, which requires a flexibility in the choice of
resolving power.
\paragraph{5.}  It is desirable to improve the spatial resolution of
missions like $ISO$ and $Spitzer$ in the same spectral range, which had
telescope sizes respectively of 60 and 85\,cm. A diameter
of 1.2\,m would give a diffraction limit of 2\rq\rq\ (FWHM) at 9.7\,$\mu$m,
improving the contrast in the detection of compact objects against their
environment and against the parasitic background.

\section{The $H2EX$ instrumental solution}
\label{h2ex_solution}
  
Imaging spectroscopy in the infrared can be achieved using different
techniques \cite{maillard_pasp}, a long slit grating spectrometer with
or without an image slicer (integral field unit), a Fabry-Perot (FP) and
an imaging Fourier transform Spectrometer (IFTS).  The possible
combination of field coverage, spectral resolution and spectral coverage
with respect to the scientific requirements oriented the final choice
towards the Imaging FTS as offering the following advantages, of
interest for $H2EX$:\\ 
- \textbf{direct imaging} of a wide field,
yielding the best image quality (no image reconstruction), which also
relaxes the pointing tolerances compared with the slit spectrometer used
in mapping mode,\\ 
- \textbf{high spectral resolution} on a wide
field. This point made a decisive element in the choice,\\ 
- \textbf{large multichannel advantage} by the simultaneous observation of
a huge number of spectra,\\ 
- \textbf{complete flexibility} in the
choice of spectral resolution, from a few hundreds up to several tens of
thousand, within the range determined by the maximum attainable optical
path difference (OPD),\\ 
- \textbf{large spectral coverage} with a
single instrument, important advantage for a mission as $H2EX$,\\ 
- \textbf{absolute wavelength calibration} of all the spectra from the
reference laser line used to control the OPD.\\

 As for any spectroscopic instrument, space gives full access to regions
totally blocked or strongly perturbed by telluric absorptions in the 8 to
29\,$\mu$m range. The other fundamental benefit is the extremely low
thermal background, particularly for a mid-infrared instrument, obtained
with the instrument being placed at L2. This condition is of major
importance for a FTS, which, by its multiplex properties, is limited in
sensitivity by the thermal background integrated over the full bandpass of
the entrance filter.  This often leads to conclude to a \textit{multiplex
disadvantage} of the FTS. The favorable background conditions in space
and the large multichannel advantage from the wide field make
possible an optimum sensitivity (Sect.~\ref{sensitivity}). Other advantage
of space compared to ground for an imaging instrument, is the absence of
image smearing, and of scintillation noise in the intensity measurements
due to turbulence. As a consequence, the image resolution can be limited by
the telescope diffraction, and the spectral instrumental lineshape be the
theoretical sinc function, purely defined by the reached OPD. Last, space
enables long observing time of the same field, important for deep surveys
at high spectral resolution.

\section{Description and key characteristics of the $H2EX$ payload}
\label{h2ex_key}
The $H2EX$ payload is composed of a telescope and a wide-field Imaging
FTS, adapted to cover the 8 to 29\,$\mu$m range.  

   \subsection{General optical design}
   \label{payload}

The IFTS is directly inspired from $BEAR$, the prototype of astronomical
IFTS in the near infrared, built for the CFH telescope
\cite{maillard00}, in operation until 2002. It worked at a high spectral
resolution of about 30\,000 in the K band, but with a field of view of
only 25\rq\rq\ diameter.  For $H2EX$ the matching to a much wider field
with the same resolution in the mid-infrared, required a completely new
design.  An adaptation of the platform developed for the $Planck$
mission proposed to reduce the project risks and cost, and simplify
the qualification, integration and test sequences defined
the allowed volume and mass.  The optical design  was  driven
by the final image quality across the field, with as a goal, to be close to
the telescope diffraction limit for the shortest H$_2$ wavelength at
9.7\,$\mu$m.   The FTS is a dual output interferometer with a cat's eye
mirror combination in each arm, which are afocal, off-axis reflective
systems made of a large concave mirror and a small convex mirror, instead of
cube corners traditionnally used.  The cat's eye takes more volume than the
cube corner, but has the specific advantage of making possible a pupil
imagery within the interferometer, mandatory for a wide-field instrument.
Each cat's eye configured as an Offner relay, reimages 1:1 the entrance
pupil made on the beam splitter by the foreoptics, on the beam
recombiner. Other interesting property, the secondary mirror is small and
can be fastly translated for an active correction of the OPD without moving
the entire assembly, which makes possible a high bandpass servo-control
system. This property, put in operation on $BEAR$, is proposed to be used
in the control of the OPD for $H2EX$ (Sect.~\ref{opd_control}).

The complete optical combination consists of an afocal Ritchey-Chr\'etien telescope
followed by an Offner relay to provide the parallel beam to the
interferometer and make a 1:1 image of the telescope secondary mirror, the
entrance pupil, on the beam splitter.  The  primary mirror has a fast f/ratio
(f/0.75) with a diameter of 1.32\,m (central
hole diameter: 0.21\,m) to accept a 20\rq$\times$20\rq\/ field of view for
a useful beam of 1.20\,m, to form a compact telescope (distance
primary -- secondary: 933\,mm). The secondary mirror serves as the entrance pupil
in order to minimize the thermal background from the primary mirror.  
Finally, the main parameters of the $H2EX$ payload
are summarized in Table~\ref{tab:param}.

\begin{table*}[!ht]
	\caption{Telescope and Imaging FTS parameters}
	\centering
		\begin{tabular}{l| r}
\hline\noalign{\smallskip} 
Telescope primary mirror & \hfill{1.32 m~~~}\\ 
Primary focal ratio & \hfill{f/0.757~~~}\\ 
Useful telescope diameter & \hfill{1.20 m~~~}\\
Secondary mirror diameter & \hfill{(entrance pupil) 80 mm~~~}\\ 
Entrance field-of-view &\hfill{20\rq$\times$20\rq~~~}\\ 
Airy disk radius &\hfill{from 2.0\rq\rq at 9.7\,$\mu$m to 5.9\rq\rq\ at 28.2\,$\mu$m~~~}\\
%\hline
\tableheadseprule\noalign{\smallskip} 
Interferometer &\hfill{dual output with cat's eye systems~~~}\\ 
Beam splitter diameter &\hfill{80 mm~~~}\\ 
Beam splitter coverage &\hfill{8\,--\,29\,$\mu$m~~~}\\ 
Maximum carriage travel & \hfill{50 mm~~~}\\ 
Maximum resolution &\hfill{0.032\,cm$^{-1} \Rightarrow$ 32\,000 at 9.7 $\mu$m~~~}\\ 
%\hline
\tableheadseprule\noalign{\smallskip} 
Imaging collimator &\hfill{3-off axis mirror system~~~}\\ 
Output image quality &\hfill{Strehl ratio $>$\,0.6 at 10\,$\mu$m on a 10\rq\ field radius~~~}\\ 
Detectors &\hfill{two 1K$\times$1K Si:As IBC/25\,$\mu$m pitch~~~}\\ 
Image sampling & \hfill{1.2\rq\rq\/$\times$1.2\rq\rq /pixel~~~}\\ 
Filters narrow (2\%) & \hfill{S(0), S(1), S(2), S(3)~~~} \\ 
~~~~--~~~ broad ($\sim$\,25\%)&\hfill{centered at 9.4, 13.0, 17.4,
 22.2\,$\mu$m~~~} \\ 
~~~~--~~~ wide &\hfill{8 - 25\,$\mu$m~~~} \\
\noalign{\smallskip}\hline 
\end{tabular}
	\label{tab:param}
\end{table*}  

   \subsection{Interferometer}
\begin{figure}[!ht]% fig. 4
\centering
\includegraphics[height=0.90\textwidth,angle=-90,clip=true]{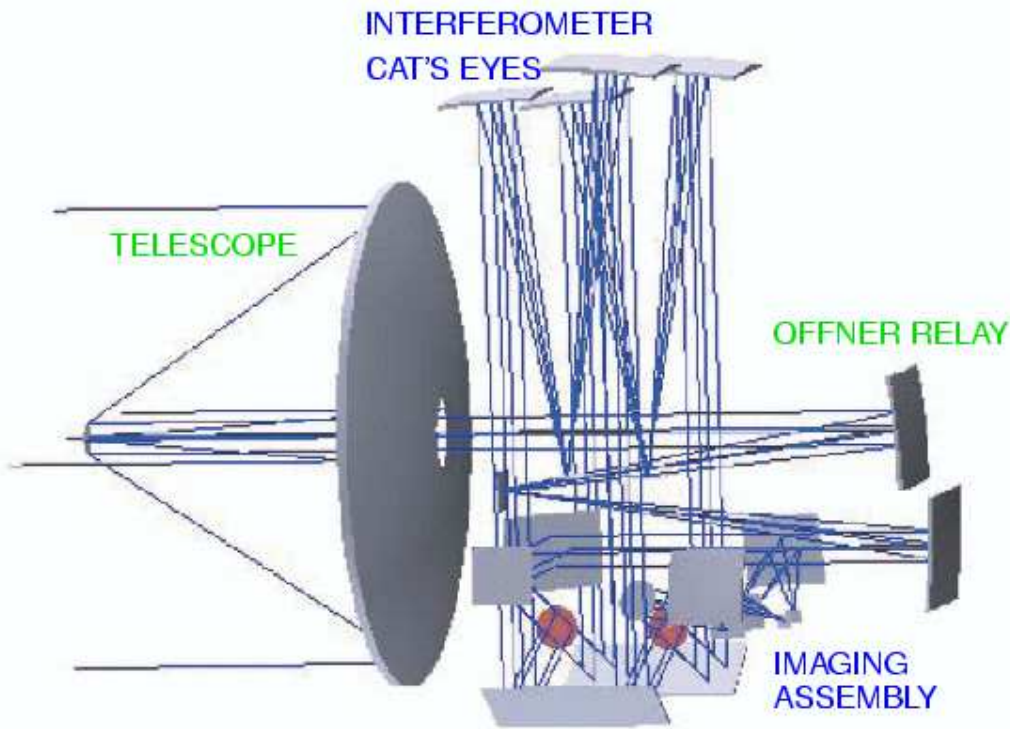}
%/tmp_mnt/netapp/users_home2/maillard/instrum/H2EX/paper_PASP/slide11_papH2EX.ps}
\caption{  
General optical layout of the $H2EX$ payload with the main
  sub-assemblies: afocal telescope, Offner relay, interferometer with the
  two parallel cat's eyes, imaging assembly with a 3-off axis mirror wide
  field corrector on each output beam.  The 1.32\,m telescope diameter and
  the 2.2\,m distance between the telescope secondary and the back of the
  Offner relay define the overall size of the payload.}
\label{fig:overview} 
\end{figure}

The entrance relay and the interferometer cat's eyes are made of spherical
mirrors with the same geometry. The important point is that through the two
folding mirrors following the entrance relay, the planes which contains the
input and the output beams in each system are strictly perpendicular. The
succession of two Offner relays with the appropriate orientation of a
system with respect to the other, perfectly compensates by their
combination the astigmatism introduced by a cat's eye on a wide field.
Actually, it was not possible to strictly keep this ideal compensation. The
length of the entrance relay had to be slightly increased (primary mirror
radius from 2000 to 2200\,mm in order to provide more room for the
telescope support) without lengthening as well the cat's eyes because of
the overall volume constraints. A diameter of 80\,mm for the parallel beam
in the interferometer has been chosen. With this beam diameter the
theoretical fringe contrast is $\geq$\,97\% across the whole field for
R\,=\,30\,000 at 9.7\,$\mu$m. On the other hand, this diameter, a little
bigger than it would have been necessary, was chosen to help correcting the
final image quality by limiting the angular magnification of the field size
to 1200/80\,=\,15, which means a maximum angle of incidence of
2$^{\circ}$\,30\rq\ in the interferometer for the 10\rq\ field radius.\\
The maximum OPD has been set to 187\,mm to reach a resolving power
(R~=~$\sigma$/d$\sigma_{\rm FWHM}$) up to 11\,000 at S(0)
($\sigma$~=~354.6\,cm$^{-1}$), which corresponds to a resolution of $\sim$\,32\,000 at
S(3) ($\sigma$~=~1035\,cm$^{-1}$). In the optical design of the
interferometer the two arms are folded by a flat mirror to bring them
parallel. With this design the two cat's eyes can be installed in the same
carriage for an OPD change by a push-pull motion of the two arms
(Sect.~\ref{opd_control}). The adopted configuration makes the overall size
of the payload relatively compact to fit into the $Planck$ platform. 
The final resulting optical layout of the $H2EX$ payload is presented in
Fig.~\ref{fig:overview}.

    \subsection{Beam splitters}    
  The number of non-hygroscopic, transparent optical materials over the
entire 8 to 29\,$\mu$m, with the minimum absorption, is very limited. Two
materials are available, which can be produced with the required size
and which are not hygroscopic, KRS-5 and diamond. They suppose to find the
adapted multilayer coating on one side and the AR coating on the opposite
side on this very broad spectral range. Hence, a completely different
solution is considered, which consists of a beam splitter made from a
3-micron polypropylene film with a multilayer coating. This material has
very good transmission properties beyond 8\,$\mu$m and can easily make a
80\,mm plate with the appropriate optical quality. This solution has been
developed for a spaceborne FTS built by NASA Langley Res. Center, to
measure the tropospheric spectrum between 10 and 100\,$\mu$m
\cite{first}. A related technique developed at Cardiff University \cite{ade}
with the same substrate, is based on metal mesh filters, and is mounted on
the $Herschel/SPIRE$ FTS.  These two solutions must be tested to get efficient
beam splitters down to 8\,$\mu$m. The realization of the beam splitters is
 a critical issue of the interferometer, but has possible solutions.

   \subsection{Imaging assembly}
\begin{figure}[!ht]% fig. 5
\centering
\includegraphics[width=0.75\textwidth,clip=true]{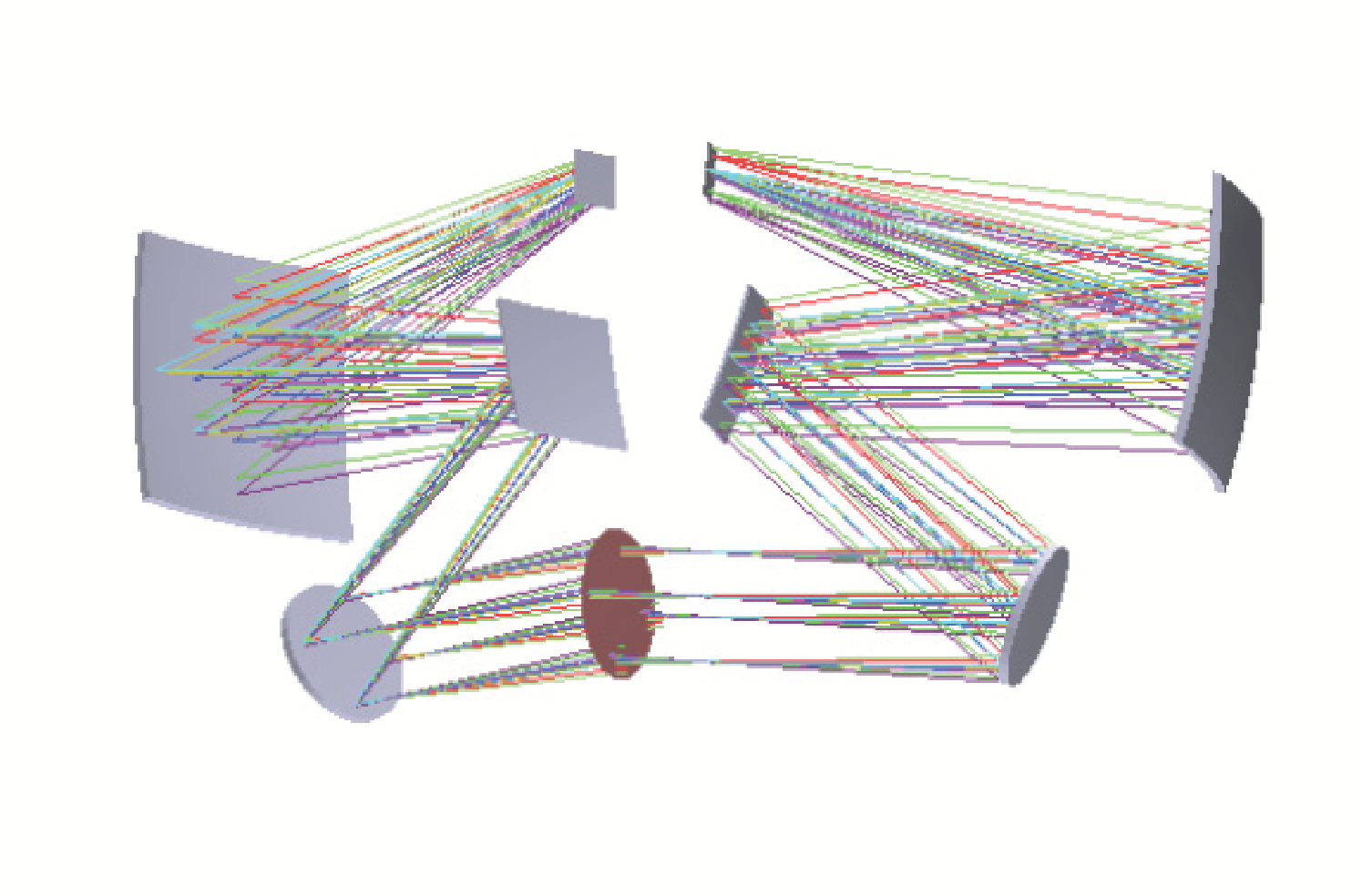}
%/home/mure1/maillard/instrum/H2EX/20060206b.ps}
\vspace{-1.0truecm}
\caption{ Imaging assembly. After the output beam splitter
  (plate in the middle of the graph) a 3-off axis mirror wide-field
  corrector is placed on the reflected and transmitted beams from the two
  arms of the interferometer, to image the entrance field on a 1K$\times$1K Si:As
  detector (25\,$\mu$m pitch). The common axis to the three mirrors is
  perpendicular to the detector.}
  \label{fig:3M_config}
\end{figure}

\begin{figure}[!ht]% fig. 6
\centering
\includegraphics[height=0.75\textwidth,angle=-90,clip=true]{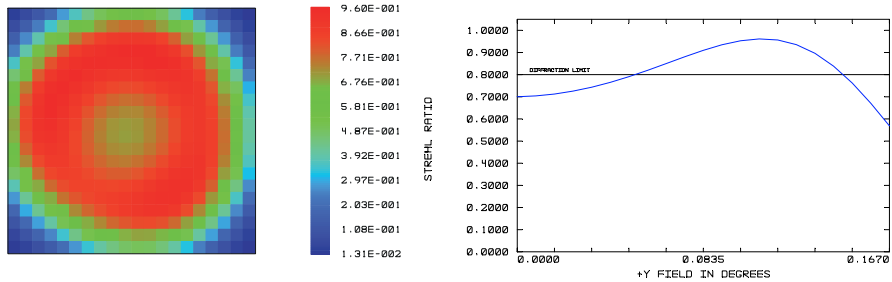}
%/tmp_mnt/netapp/users_home2/maillard/instrum/H2EX/paper_PASP/slide7_papH2EX.ps}
\caption{ \textit{Left:} Strehl ratio map at 10\,$\mu$m over
  the 20\rq$\times$20\rq\ field obtained with the full optical combination:
  telescope, Offner relay, one interferometer arm and 3-off axis mirror
  imaging assembly (Fig.~\ref{fig:3M_config}).  \textit{Right:} Cut of the
  map along the $+$ Y axis.}
  \label{fig:SR_map}
\end{figure}

 The chosen configuration for each output beam to image the field on a
 1K$\times$1K detector with 25\,microns pitch is a 3-off axis mirror
 collimator (concave -- convex -- concave), to make a compact system, with
 the three mirrors having a common axis (Fig.~\ref{fig:3M_config}). The
 pixel size requires a f/4 collimator. The optimization is made with
 aspheric surfaces. The two detector arrays are mounted back-to-back on
 the two sides of the same cold head at 5\,K. However, this collimating
 system is limiting the final image quality, which is estimated by the
 Strehl ratio map shown in Fig.~\ref{fig:SR_map}. At 10\,$\mu$m the Strehl
 ratio is comprised between 0.95 and 0.56 over a 10\rq\ radius field. At
 the extreme corners of the field for this wavelength the Strehl
 ratio goes down to 0.013 in the worst corner, the astigmatism introduced
 by the 3-mirror assembly becoming strongly prominent. The Strehl ratio map
 clearly shows the usable field.

A filter wheel cooled at the same temperature as the detectors, placed
between the third mirror (Fig.~\ref{fig:3M_config}) and the detector
makes possible to select the spectral domain of the observation. As
indicated in Table~\ref{tab:param} a basic set of 9 filters is mounted,
with four 2\%-band filters, centered respectively on each H$_2$
line. Each of them besides S(0), also represents a specific $z$ value
for a shorter wavelength H$_2$ line. For example, the S(1) filter
corresponds to the S(5) line at z\,$\simeq$\,1.5. A fifth narrow-band
filter can be added, for example at 19.4\,$\mu$m to give access to the
0\,-\,0 R(5) line of HD. Four broad filters are indicated
$\sim$\,25\%-band in the table to cover the 8\,-\,25\,$\mu$m domain with
adjacent bandpasses. The set is completed by a 8\,-\,25\,$\mu$m
wide-band filter. The longest wavelength of these broad and wide filters
is limited to $\sim$\,25\,$\mu$m to avoid entering the region where the
thermal background becomes too high (Fig.~\ref{fig:background}).

   \subsection{Post-submission optical design}
\label{post_subm}
Since the submission of the $H2EX$ proposal at the end of June 2007 the
optical design has been slightly revised. The current layout implies an
afocal telescope. This choice makes the instrument not transposable to a
classical Ritchey-Chr\'etien telescope for further applications. In
addition, the telescope secondary mirror diameter is small (equal to the
beam splitter diameter) which leads to severe constraints on
the mirror positioning (see Sect.~\ref{h2ex_meca}). A different optical
design has been tested \cite{maillard_pasp} with a f/10 Ritchey-Chr\'etien telescope,
which keeps the same optical design for the interferometer, a
comparable  overall size of the payload, and with an equal final image
quality while removing all the tight tolerances on the telescope
building. An aspheric field mirror and an off-axis collimator replace the entrance
Offner relay.    

\section{$H2EX$ mechanical design}
\label{h2ex_meca} 
\begin{figure}[!h]% fig. 7 
%\centering
\includegraphics[width=0.60\textwidth,clip=true]{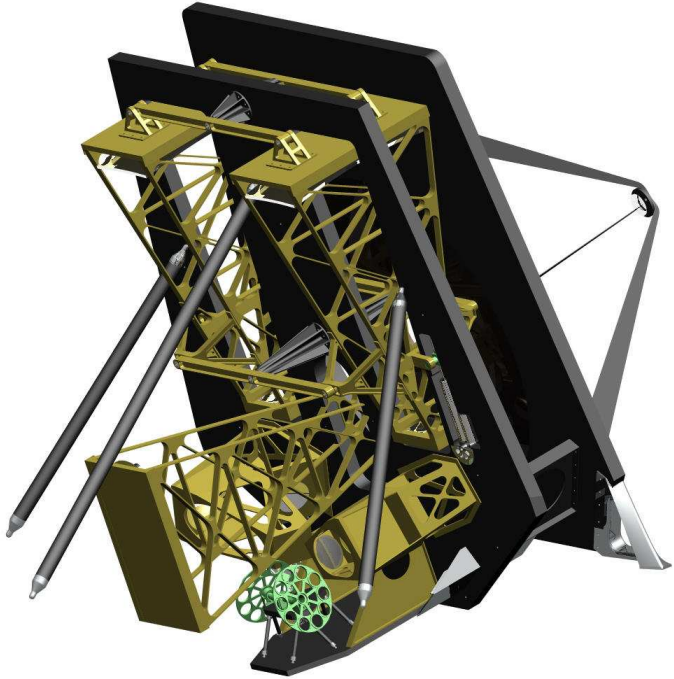}
%/home/mure1/maillard/instrum/H2EX/proposal_CV/Vue_meca_all.eps}

\vspace{-4.50truecm}
\hspace{0.65\textwidth} 
\begin{minipage}[l]{0.35\textwidth}
\caption{View of the instrument at the back of the telescope with the 3
   legs to mount the payload on the platform, showing the main bench with
   the Offner relay and the two cat's eyes. The baffle which thermally
   isolates the 5K part is removed to display the imaging assembly with the
   filter wheel at each focal plane. An external baffle to improve the
   passive cooling of the payload  covers the whole instrument.}
\label{fig:struct_design}
   \end{minipage}
 \end{figure}

A first order instrument layout has been studied
around the optical configuration (Fig.~\ref{fig:overview}) which is
shown in Fig.~\ref{fig:struct_design}.  The telescope and the IFTS, each
of them supported by a honeycomb bench, are designed to be assembled and
tested separately.  The IFTS itself is made of the following
sub-systems: entrance Offner relay, cat's eyes, imaging assembly, which
can be aligned separately.\\ 
All the mirrors for the telescope and the
IFTS are fabricated in lightened C/SiC (surface density
15\,kg/m$^2$). This composite material presents a high stiffness, a low
coefficient of thermal expansion (2.2\,10$^{-6}$/K at room temperature,
around 0 at 100\,K) and a high coefficient of thermal conductivity.
Several space mirrors ($Herschel$, $Aladin$, $Gaia$) have already been
fabricated with this material, the biggest one being the 3.5-m
$Herschel$ primary mirror made by the combination of 12 pieces.  The
1.3-m $H2EX$ mirror can be made in a single piece.  The secondary mirror
of the telescope is linked to the primary by a tripod made of carbon
fiber with narrow fins (5\,mm thickness) fixed at the back of the
secondary, also to limit the thermal background from the unit.  Due to
the fast f/ratio of the telescope primary mirror (f/0.75), the centering
and the focus of the secondary are critical.  Its distance to the primary must be
precise as ${\rm d} = 930 \pm 0.025$\,mm. The contraction of the carbon
fibers from room temperature to 40\,K is of the order of
370\,$\mu$m. Then, the focusing of the telescope requires a motorized
micrometric focusing mechanism at the back of the secondary, activated
on ground for the test operations at room and cryogenic temperature.
Before launch the secondary should be blocked for the best focus
determined at 40\,K. However, this constraint can be cancelled by a
different optical design of interferometer foreoptics (see Sect.~\ref{post_subm}).

    \subsection{OPD control mechanism}
\label{opd_control}
\begin{figure}[!ht] % fig. 8

\includegraphics[scale=0.45]{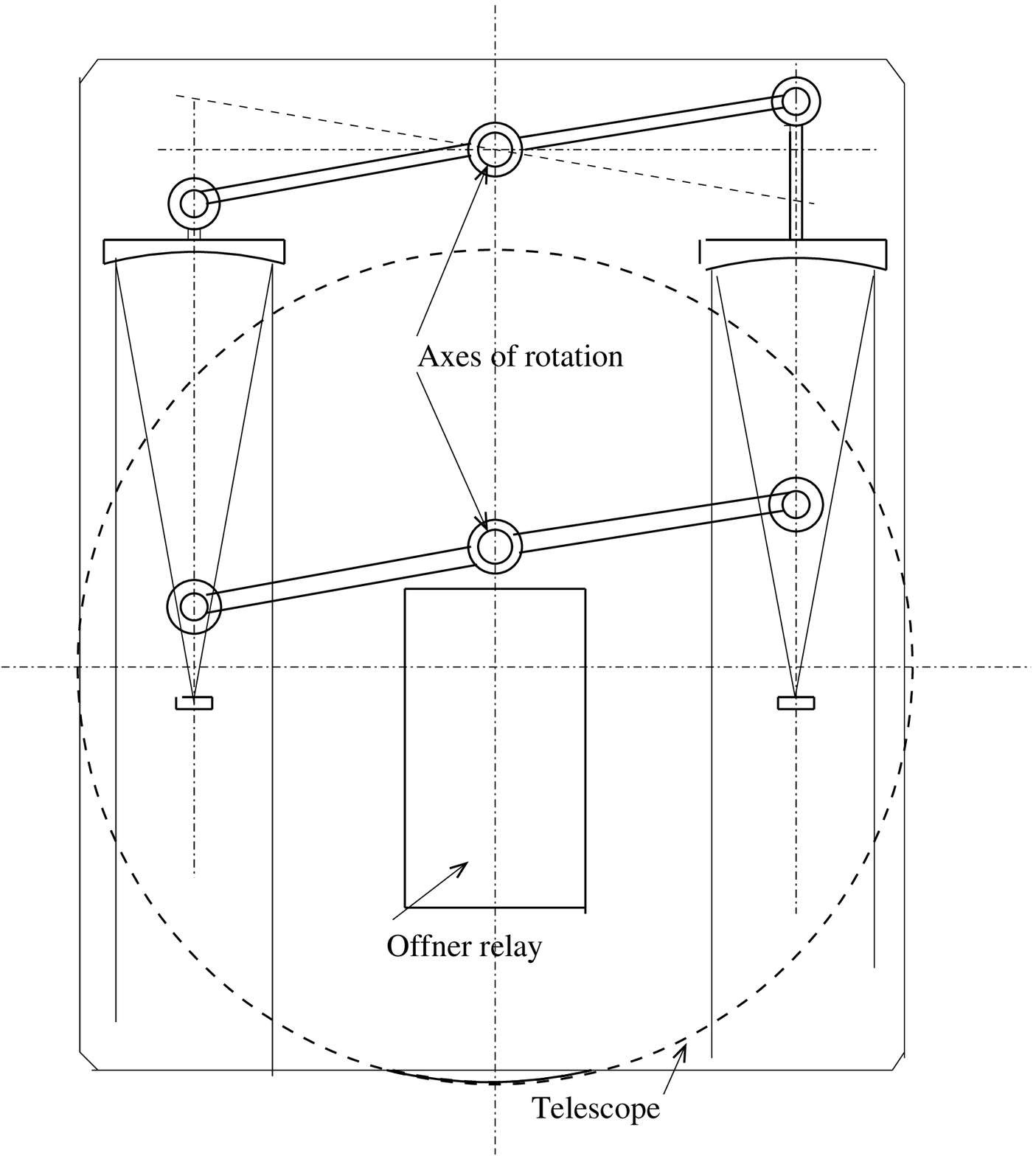}  
%/home/mure1/maillard/instrum/H2EX/H2EX_OPD2.eps}
\includegraphics[scale=0.45]{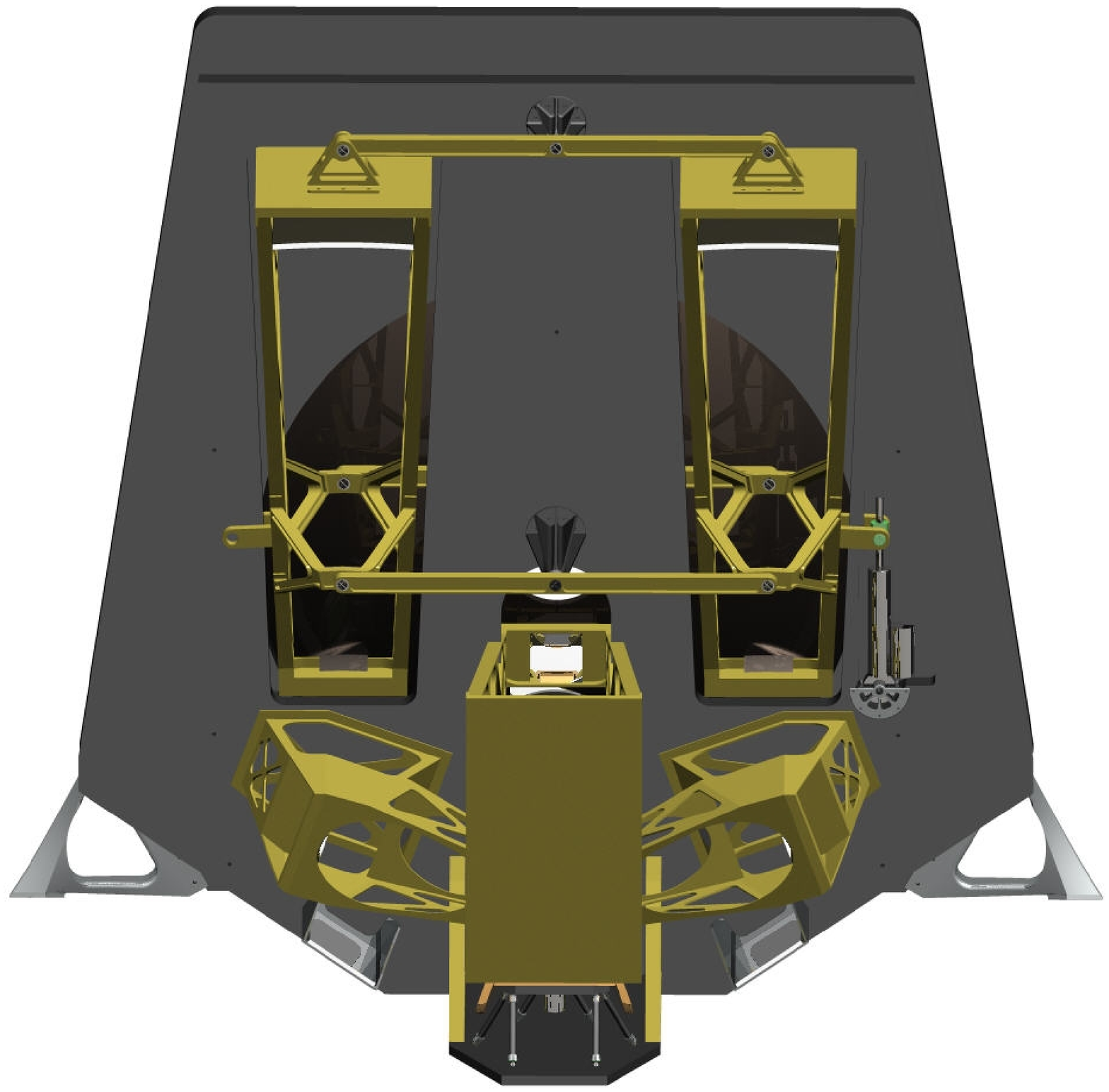}
%/home/mure1/maillard/instrum/H2EX/proposal_CV/Vue_meca_OPD.eps}
\caption{\textit{Left:} Principle of the carriage motion with the two cat's
eyes in push-pull displacement as the parallel sides of a
parallelogram. The axes of rotation are frictionless, made of flexural
pivots. The distance between the two cat's eyes is equal to
2$\times$430\,mm. For a travel of 50\,mm, maximum transverse displacement
of 0.7 mm.\hfill\break \textit{Right:} View of the OPD scanning mechanism
mounted on a honeycomb bench. At the right side, the endless screw (20\,mm
diameter, pitch 0.5\,mm) to move the parallelogram.}
	\label{fig:scan_opd}
\end{figure}

The OPD is scanned in a \textit{step and look} mode which is the heart of
the instrument.  The two cat's eyes are mounted as the two parallel sides
of a parallelogram (Fig.~\ref{fig:scan_opd}), making possible a push-pull
motion which limits the required mechanical course of the carriage to $L$
for a maximum OPD variation equal to $4\times L$. In practice, the maximum
travel of each cat's eye is equal to 50\,mm. It does not correspond exactly
to a maximum OPD of 200\,mm, but to 187\,mm, since the scans start a little
before the zero-OPD to fully cross the main burst of fringes at zero-OPD.
 Flexural pivots provide frictionless axes.  The
whole system is perfectly balanced for the minimum reaction on the pointing
of the spacecraft. The step-by-step linear displacement of the cat's eyes
is realized from the rotation of a space-qualified endless screw with a
0.5\,mm pitch, of 20\,mm diameter, mounted in parallel to the optical axis
of the cat's eyes, and a nut with planetary rollers.  The motor is
activated at each step and is cut during the constant 20\,s integration
time per image which follows the motion, making negligible the heat load.
The distance between the two cat's eyes produces a very small lateral
motion (0.7\,mm) for the maximum travel.  The rms error on the step size,
which must be $\leq 10$ nm, is reached by a servo-control of the OPD.  The
light secondary mirror (90$\times$90\,mm, weight 120\,g) of each cat's eye
is mounted on a piezoelectric actuator to provide the needed correction to
the displacement given by the step-by-step motor which has a resolution
close to 1 micron. As such, the motion of the piezo-actuators is always
very small (within 5 to 10 microns) and only requires a low voltage. The
piezo-actuators are driven by the real-time measurement of the OPD provided
by a near-infrared stabilized laser diode whose beam, injected by optical
fiber, is parallel to the science beam coming from the telescope. Such
diodes are currently used on several FTSs in space. This type of mechanical
displacement and its coupling to piezo-actuators has been in operation for
years on the $BEAR$ Imaging FTS.
 
    \subsection{Thermal design}
 The cooling of the payload is an important aspect of the performance of
 the instrument.  The design which is chosen is a simplified copy
of the architecture developed for the $Planck$ mission since the coolest
elements are two Si:As arrays and the two filter wheels, operating at
$\sim 5\,$K. This part is protected from the 40\,K emission by a baffle cooled
at 18\,K. The cooling of the detector box can be made by a JT/Stirling
cooler combination as used for MIRI. The 35\,K goal can be
 reached by strictly limiting the heat load of the entire instrument.

   \subsection{Technology readiness level (TRL)}
Several astronomical FTSs have already been put in operation in space. The
most famous of them on the Voyager spacecrafts, $IRIS$, returned
mid-infrared spectra all along the visit of the four giant planets, from
1979 to 1986. There are currently in space, $PFS$ on Mars Express, $CIRS$
on Cassini, $FIS$ on Akari. $SPIRE$ on Herschel will be launched in
2008.  Many other FTSs have been developed for remote sensing of the Earth
atmosphere. The most recent, $IASI$, developed by CNES, which reaches a
maximum resolution of 0.35\,cm$^{-1}$, was launched successfully
in October 2006.  Imaging in the 3.6 to 15.5\,$\mu$m range is obtained by a
scanning the field with 4 single-pixel detectors as the platform moves along its 
orbit around the Earth. A new generation instrument for
the same spectral domain, $GIFTS$ (Geosynchronous Imaging FTS), part of
NASA Earth Observing-3 program, already tested in balloon, is a direct
imager as $H2EX$.

\begin{table}[!ht]
\caption{Main $H2EX$ sub-systems with their TRL rated from ESA code}
        \centering
                \begin{tabular}{l| l| l}
\hline\noalign{\smallskip}
Sub-system & Heritage & TRL \\
\tableheadseprule\noalign{\smallskip}
C/SiC Telescope & Herschel, Gaia & 8\\

OPD control system & BEAR, GIFTS & 6, 7\\

Beam splitter & FIRST, SPIRE & 9\\

diode laser & IASI, GIFTS & 9 \\

Filter wheels& ISO, MIRI  & 9, 8  \\

Si:As detector & MIRI, WISE & 8 \\

Cryogenics & Planck, MIRI & 8, 6 \\

Data compression& Mars Express& 9\\
\noalign{\smallskip}\hline
                \end{tabular}
        \label{table:legacy}
\end{table}

 The $BEAR$ instrument has given a solid
experience on the type of data which will be acquired by $H2EX$, with the
development of a complete chain of programs, from the raw data to the
calibrated spectral cube, which has just to be adapted to the $H2EX$
fields. The $H2EX$ challenge consists of building an Imaging FTS with a
similar resolution but with a much wider field and to bring it to
space. Other aspects as filter wheels, infrared detectors, cryo-generators,
are common to other infrared space missions. From all this experience, in
Table~\ref{table:legacy} are listed the main $H2EX$ sub-systems, showing
that for all of them their TRL is $\geq 6$.

\section{$H2EX$ sensitivity}
\label{sensitivity}

From the design of the $H2EX$ payload a careful analysis of the
 the minimum detectable flux has been conducted.  For
sufficiently long integrations, the limit of sensitivity is dominated by
the background photon noise.  The parasitic flux in space is the sum of the
zodiacal light and the thermal emission from the payload at L2
(Fig.~\ref{fig:background}).  The entrance field is imaged on two detector
arrays, the two complementary outputs of the interferometer.  To obtain the
final sensitivity, the noise on each pixel of the detector area over which
the flux is integrated is added quadratically and the signal is the sum of
the two complementary output flux. The optical throughput and the optics
emissivity for each output beam used for the sensitivity calculations,
taking into account all the optical elements (see \cite{maillard_pasp} for
details), are given in Table~\ref{tab:detspecs}. 
Two observing cases have been considered: extended gaseous regions and
 point sources. The main results are
presented in Table~\ref{mode_obs} for six combinations of
filter bandwidth and resolution, which have been defined for the $H2EX$
surveys (Table~\ref{tab:surveys}).

\begin{table}[!ht]
\caption{Instrument and detectors inputs used in sensitivity computations}
\centering
\begin{tabular}{l|l}
\hline\noalign{\smallskip}
Optics temperature   & 40\, K \\
Filter temperature   &  5 - 6\,K\\
Mirror reflectivity  & $R\,=\,0.99$ (gold-coated)\\
Beam splitter transmission & $T_S\,=\,0.52$ \\
~~~~~~~--~~~~~~~~~ reflection & $R_S\,=\,0.45$ \\
~~~~~~~--~~~~~~~~~ absorption & $A\,=\,0.03$ \\
Optical throughput   & $\Phi_{I}\,=\, \Phi_{II}$\,=\, 0.39$^{(1)}$   \\
Optics emissivity    & $E_{I}\,=\,E_{II}$\,=\, 0.17$^{(1)}$ \\
Peak filter transmission & 0.8 \\
Detector  QE         & 0.5 for $8 < \lambda < 25\,\mu$m, 0.13$^{(2)}$ at 28.2\,$\mu$m \\
Well capacity        & 2 $\times 10^5{\rm  e^-}$ \\
Readout noise        & 20 ${\rm e^-}^{(3)}$ \\ 
Dark current         & 10 ${\rm e^-/sec}$ \\ 
\noalign{\smallskip}\hline
\multicolumn{2}{l} {$(1)$: same results for $T_S\,=\,45\%$ and
  $R_S\,=\,52\%$}\\
\multicolumn{2}{l} {$(2)$: with AR coating peaked at 28\,$\mu$m}\\
\multicolumn{2}{l} {$(3)$: by measuring fluxes with a 2.7\,s frame time}\\  %with Fowler-8 sampling\\
 \end{tabular}
\label{tab:detspecs}
\end{table}
  
 For high spectral resolution work the observing strategy
 consists of using high-peak transmission 2\% narrow-band filters to
 isolate each observed H$_2$ emission line.  The sky background flux
 from space and the thermal emission from the low-emissivity cold
 instrument (Fig.~\ref{fig:background}), within the narrow bandpass,
 form the incoming flux which limits the sensitivity. Since emission
 lines are observed, the signal-to-noise ratio for the peak of the
 emission line is independent of the resolution, as long as the line is
 not resolved. Finally, the line contrast is improved by increasing the
 resolution. \textbf{These three conditions: 1) narrow-band filter in
 low-background environment, 2) lines in emission, 3) high spectral
 resolution, provide the optimum sensitivity to the $H2EX$ Imaging FTS}.

\begin{table*}[!ht]
	\centering
\caption{ Definition and sensitivity of the $H2EX$ observing modes}
\begin{tabular}{@{}c| @{}c  r r l @{}c @{}l l}
\hline\noalign{\smallskip}
 Mode &Filter  & $R$~~ & Obs.~~  &~~Targets & Tracer 
                              &D$_1$$^{(1)}$/{\bf Goal}$^{(2)}$ & D$_2$$^{(1)}$ \\
     & width  &        & time (h)  &          &      
                              &                                 &  \\
\tableheadseprule\noalign{\smallskip}
  1  & 2\%      & 11\,000   &  24 & star form. regions   & H$_2$ 
                                &~5.0$^{(3)}$/{\bf 1.3}$^{(3)}$ & 1.5$^{(3)}$\\
     &           &           &     & nearby galaxies& H$_2$ &  & \\
\hline\noalign{\smallskip}
  2  & 2\%    & $\geq$ 11\,000&100&star form. regions & H$_2$
                                        &~29/{\bf 6.8}  &3.5 \\
     &          &           &    &circumstel. disks& H$_2$ &  & \\ 
\hline\noalign{\smallskip}
  3  & 2\%    & 1000      & 100 & distant galaxies& H$_2$ 
                                        &  &1.3 \\
\hline\noalign{\smallskip}
  4  & 25\%     & 1000      & 100 & distant galaxies& ions 
                                        &~8.1/{\bf 4.5}  & 4.3 \\
     &           &           &     &                & H$_2$&   & \\
\hline\noalign{\smallskip}
  5  & 25\%     & 10000     & 24  & Mol. Clouds  & molecules
                                        &~16/{\bf 9.0} & 8.6 \\
\hline\noalign{\smallskip}
  6  & 8 - 25\,$\mu$m& 100    & 36 & distant galaxies  &  dust 
                            &\multicolumn{2}{c}{60/{\bf 36} $\mu$Jy} \\
\hline\noalign{\smallskip}
\multicolumn{8}{l} {$(1)$:  5\,$\sigma$ detectivity in
  10$^{-20}\,$W\,m$^{-2}$ D$_1$ for S(0) and the 25\% 23\,$\mu$m filter,
D$_2$ for the other filters}\\
\multicolumn{7}{l} {$(2)$: D$_1$ for passive cooling at 35\,K and QE =\,50\% beyond
 26\,$\mu$m  (Si:P detectors instead of Si:As)}\\
\multicolumn{7}{l} {$(3)$: 5\,$\sigma$ detectivity in 
10$^{-10}\,$ W\,m$^{-2}$\,sr$^{-1}$ in a 5\rq\rq$\times$ 5\rq\rq area}\\
	\end{tabular}
	\label{mode_obs}
\end{table*}

\section{The $H2EX$ spacecraft and the mission profile}
\label{spacecraft} 
\begin{figure}[!ht] % fig. 9
\includegraphics[scale=0.35,clip=true]{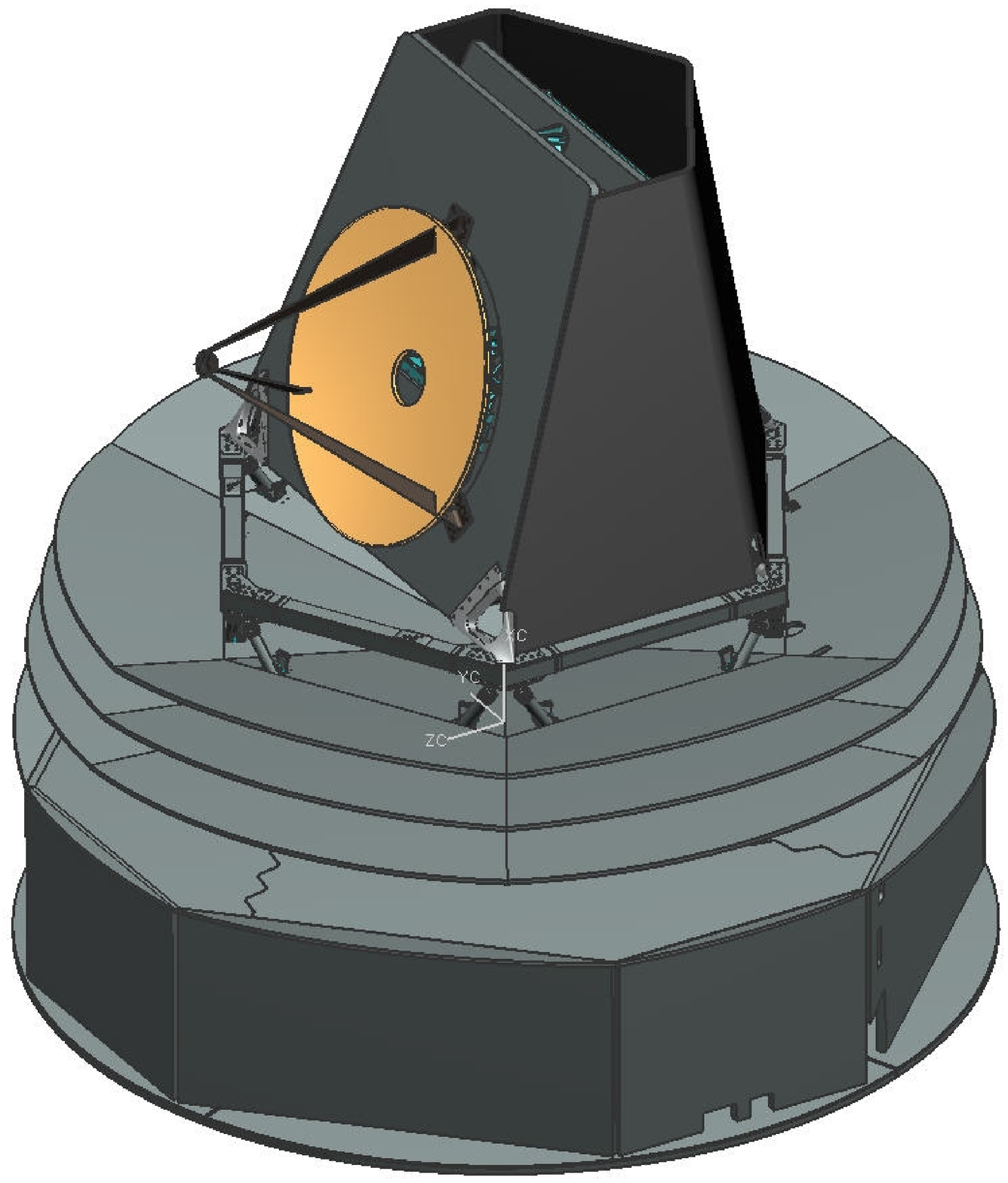}
%/tmp_mnt/netapp/users_home2/maillard/instrum/H2EX/paper_PASP/View_H2EX_new.ps}
\includegraphics[scale=0.35,clip=true]{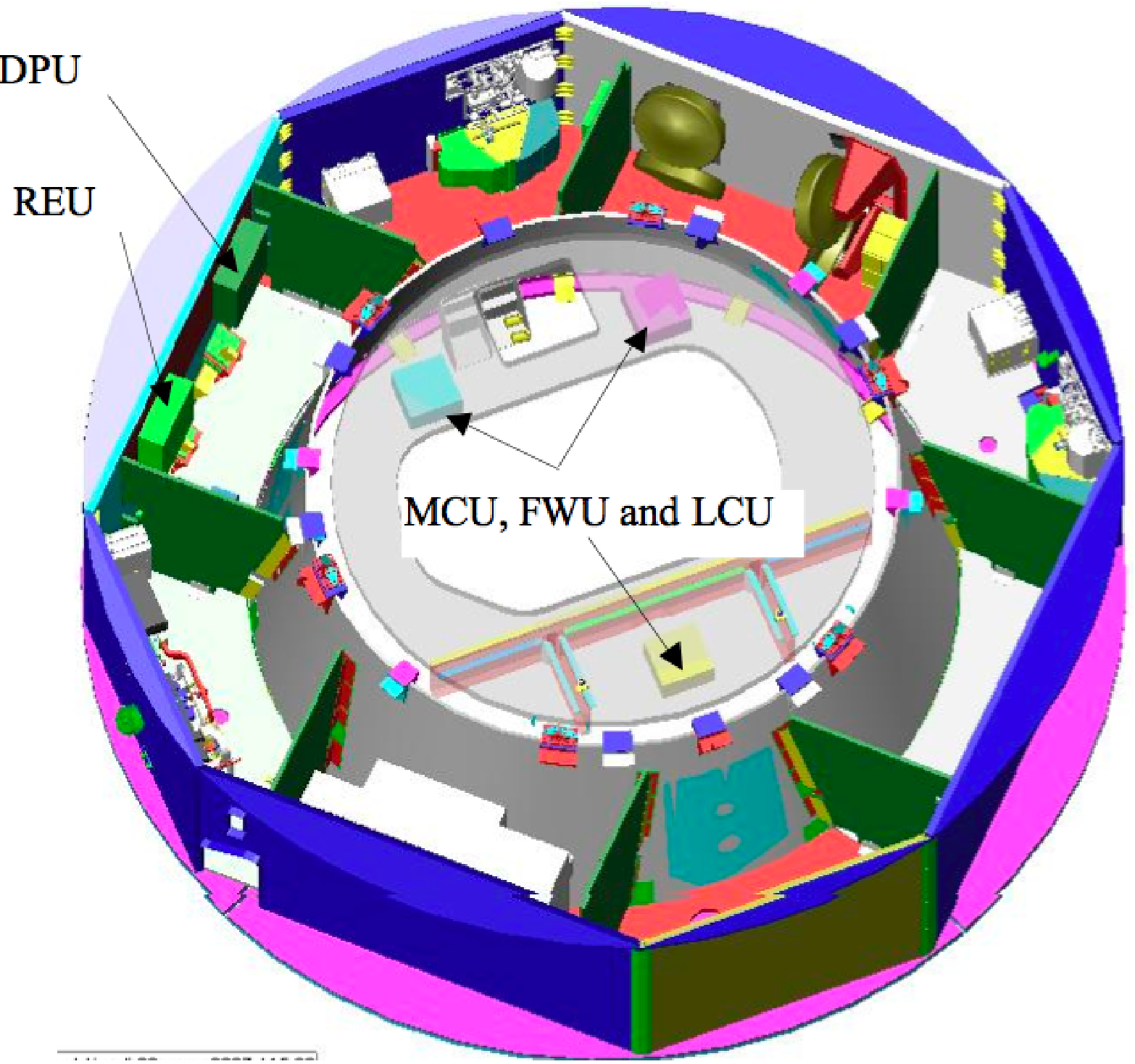}
%/home/mure1/maillard/instrum/H2EX/proposal_CV/fig_svm.ps}
\caption{\textit{Right:} View of the $H2EX$ spacecraft with the
  telescope and behind, the IFTS covered by an external shield to
  improve the passive cooling, placed on the $Planck$
  platform.\hfill\break \textit{Left:} View of the Service Module with
  in place the electronics units: the OPD Motor Control Unit (MCU), the
  Filter Wheels Unit (FWU), the Laser Control Unit (LCU), the Readout
  Unit (REU) and the Digital Processing Unit (DPU).}  \label{H2EX_craft}
\end{figure}

The $H2EX$ payload to be installed in the spacecraft
is shown in Fig.~\ref{H2EX_craft} (right) and the Service Module in
Fig.~\ref{H2EX_craft} (left).  The Service Module of the $Planck$
platform is formed by an octagonal box built around a conical tube that
supports the module cryo-structure, and with on top, an hexagonal frame
which supports the payload by three legs
(Fig.~\ref{fig:struct_design}). The $H2EX$ payload has been designed to
comply with this configuration.    The
baseline $H2EX$ cryogenic chain is derived from $Planck$.  The passive
radiator based on the 3 V-Groove insulation system meets the
40\,K requirement on the telescope and spectrometer optics, with the
hope to be able to reach 35\,K by limiting the internal heat lift. For
the detectors, one JPL H$_2$ sorption cooler ensures an intermediate
18\,K stage and two RAL He Joule-Thompson 4\,K coolers working at 50\% of their full
capacity provide the necessary cooling power. The interfaces between the
passive radiators and active coolers remain the same as for $Planck$.

%Cryogenics are needed to
%keep the detectors and the filter wheels at 5\,K.  
%The specific thermal
%requirements are met from the qualified $Planck$ cryogenic chain based on
%cryocoolers.
%, with a 5\,K stage protected from 40\,K emission by a 18\,K stage.

A shield with the solar array on the bottom of the service
module maintains the necessary shadowing cone. The orientation of the
telescope permits to access most of the sky outside a narrow region around
the ecliptic poles. A given position within the accessible sky can be
observed twice a year for a duration of at least 10 days.  The $Herschel$
qualified Attitude and Orbit Control System (AOCS) complies with the three
axis stabilized $H2EX$ pointing requirement (absolute pointing error
$\pm$\,30\rq\rq, pointing stability 0.5\rq\rq over 20\,s). This adds four
reaction wheels and one internally redundant gyroscope to the $Planck$ AOCS
sub-system.  The use of a SOYUZ-Fregat launcher instead
of an ARIANE V for $Planck$ requires a launcher adapter and  diameter
reduction: 420mm for the solar array and 40 mm for the structure. 
%The satellite fits on a Soyuz-Fregat 2B rocket. 
%The $H2EX$ mission requires a 3-axis stabilized spacecraft in orbit at L2,
%similar to that of $Planck$, for passive cooling of the payload at least at
%40 K, at 35\,K as a goal, which is required to minimize the background
%noise in the mid-infrared.  

%$H2EX$ has been proposed as an ESA medium class mission with NASA and CSA
%participation to carry pointed observations by spectro-imagery of selected
%fields in the sky.  
The  scientific program (Table~\ref{tab:surveys})
is designed for a nominal 2 years mission, which could be extended since
the lifetime is not limited by cryogenic fluids.  $H2EX$ is an explorer
type mission by its instantaneous field size but the surveys require long
(24 hours) pointing for each field. The operations are closer to those
of a space observatory like $Herschel$ than of all-sky survey missions such
as $Planck$.

\section{$H2EX$ in context}
\label{h2ex_context}   

To evaluate the scientific interest of $H2EX$ the proposed instrument
must be replaced in context of other infrared facilities also able to
observe the H$_2$ rotational lines.  With a maximum resolving power
around 3000, $MIRI$, the $JWST$ mid-infrared integral-field spectrometer
\cite{wright} will cover the full 5\,--\,29\,$\mu$m range in four
channels associated to different small fields from
3.0\rq\rq$\times$3.9\rq\rq\ to 6.7\rq\rq$\times$7.7\rq\rq.  The main
characteristics is the high resolution image sampling, going from 0.19
to 0.27\rq\rq/pixel.  On ground, $TEXES$ \cite{lacy,bitner} and its copy
$EXES$ on $SOFIA$ \cite{richterm} provide a high spectral resolution, up
to 10$^5$ on the 5\,--\,25\,$\mu$m domain. Sensitivity is limited by the
sky emission.  The spectral coverage is very narrow, of 0.5\%, and the
slit length is small, typically from 6 to 12\rq\rq\ depending on the
wavelength range.  Another mode at medium resolution (15\,000) provides
a slit of 1.5\rq\rq$\times$45\rq\rq. These two types of instrument,
$MIRI$ and $TEXES$ are designed for targeted studies of distant
galaxies, ULIRGs and AGNs, of multiple stellar systems, of circumstellar
disks, but are not suited for the observation of extended sources, or to
complete a survey of these different targets over a wide field.  The
limited spectral resolution in the case of $MIRI$ will not make possible
to access the dynamics of star forming regions.

\begin{table}[!ht] 
\caption{ $H2EX$ in context of other instrumentation for the observation of H$_2$
  0\,--\,0 S(1) 17.3\,$\mu$m}
 \centering
\begin{tabular}{l|l r r r r}
\hline\noalign{\smallskip} 
Instrument/ & Ang. Res. & R~~     & Field of & Point$^{(1)}$ & Extended$^{(2)}$ \\
~~~~~~~Telescope &FWHM (\rq\rq) & $\lambda/\delta\lambda$&view (\rq\rq$^2$)& source & emission\\ 
\tableheadseprule\noalign{\smallskip}
%\hline
H2EX            & ~~3.6 & 18\,000 & 144$\times$10$^4$  & 7 & 1.5~~ \\
%\hline
\tableheadseprule\noalign{\smallskip}
IRS-LL/Spitzer   & ~~5.0   & 80    & 756 & 38 & 6~~ \\
IRS-SH/Spitzer   & ~~5.3 & 600  & 54 & 20 & 3~~ \\
MIRI/JWST        & ~~0.7 & 3000 & 52  & 0.5 & -~~ \\
TEXES/Gemini     &~~1.0   &80\,000 &  12  & 650 & -~~ \\
TEXES/Gemini     &~~1.5  &15\,000 & 67  & 50 & -~~ \\
EXES/SOFIA       &~~1.7  & 80\,000 & 12   & 650 & -~~ \\
MIR/Spica$^{(3)}$  & ~~1.4 & 3000 & 1800 & 1.5 & 0.6~~ \\
\noalign{\smallskip}\hline
\multicolumn{6}{l}{${(1)}$ 5\,$\sigma$ sensitivity, 5\rq\rq\ spatial
  resolution in
   3\,hrs, unit 10$^{-20}$\,W\,m$^{-2}$.} \\
\multicolumn{6}{l}{${(2)}$ 5\,$\sigma$ sensitivity, 5\rq\rq\ spatial
  resolution in
   3\,hrs, unit 10$^{-10}$\,W\,m$^{-2}$\,sr$^{-1}$}\\
\multicolumn{6}{l}{${(3)}$ Courtesy of B. Swinyard}\\
\end{tabular}
  \label{tab:survey_power}
\end{table}

The uniqueness of $H2EX$ is demonstrated in Table~\ref{tab:survey_power}
where we use the S(1) rotational line to compare space and ground-based
infrared facilities.  The instantaneous field of view of $H2EX$ is more
than twenty thousand times larger than that of the high spectral
resolution $Spitzer/IRS-SH$ mode and $MIRI/JWST$. In addition, the
highest spectral resolution mode of $H2EX$ is 50 times that of the
smaller-area high resolution slits of $IRS$ and 6 times that of $MIRI$.

%The good sensitivity
%comes from these parameters related to the choice of observing
%emission lines at high resolution on a small spectral range, as discussed
%in Sect.~\ref{sensitivity}.

\section{Conclusion}
\label{future} 

$H2EX$ is devoted to the detection of the first rotational lines of
molecular hydrogen, through a series of targeted surveys, going from the
Cosmic Web to the circumstellar disks in the solar neighborhood, and the
formation of complex molecules. Five fundamental aspects of the origin
of the luminous universe can be addressed through the proposed
surveys. The instrumental novelty of $H2EX$ is to offer a flexibility in
the spectral resolution choice, from low up to high values of 30\,000,
that can be tailored to the scientific case, together with a wide field
of 20\rq$\times$20\rq.  This capability is missing in the thermal
infrared astronomical instrumentation, currently in operation and in
preparation.  Space is the most favorable place for such an instrument.

\begin{acknowledgements}
The authors greatfully acknowledge help and advices from Ren\'e Laureijs
at ESA who, as former $H2EX$ PI, was the early initiator of the new $H2EX$
mission.  The preparation of the proposal was supported by the
Centre National d'\'Etudes Spatiales (CNES).
\end{acknowledgements}

%%-----------------------------

\end{document}